\documentclass[%
nofootinbib,
nobibnotes,
 amsmath,amssymb,
 aps,
 prl,
floatfix,
twocolumn, 
titlepage
]{revtex4-1}

\usepackage[dvips]{graphicx}
\usepackage{pstricks}
\usepackage{enumerate}
\usepackage{subfig}
\usepackage{amssymb,amsmath,latexsym}
\usepackage{amsthm}
\usepackage{pstricks-add}
\usepackage{pst-plot}
\usepackage{pst-xkey}
\usepackage{multirow}
\usepackage{slashbox}
\usepackage{dsfont}
\usepackage{pifont}
\usepackage{bm}
\usepackage{textcomp}
\usepackage{microtype}
\usepackage{url}
\usepackage{units}
\usepackage{mdwlist}
\usepackage[T1]{fontenc}
\usepackage{hyperref}
\usepackage{nameref}

\newcommand{\tr}{\operatorname{tr}}

\newcommand{\id}{\mathds{1}}

\newcommand{\eff}{\textbf{f}}
\newcommand{\gee}{\textbf{g}}

\newcommand{\proj}[1]{\ensuremath{\left| #1 \rangle \langle #1\right|}} 
\newcommand{\ket}[1]{\ensuremath{\left| #1 \right>}} 

\theoremstyle{plain}

\theoremstyle{definition}

\theoremstyle{remark}

\makeatletter
\let\orgdescriptionlabel\descriptionlabel
\renewcommand*{\descriptionlabel}[1]{%
  \let\orglabel\label
  \let\label\@gobble
  \phantomsection
  \edef\@currentlabel{#1}%
  \let\label\orglabel
  \orgdescriptionlabel{#1}%
}
\makeatother

\allowdisplaybreaks

\begin{document}

\title{A violation of the uncertainty principle implies a violation of the 
second law of thermodynamics}

\author{Esther H\"anggi}
\email[]{esther@locc.la}
\affiliation{Centre for Quantum Technologies, National University of Singapore, 3 Science Drive 2, 117543 Singapore}
\author{Stephanie Wehner}
\email[]{steph@locc.la}
\affiliation{Centre for Quantum Technologies, National University of Singapore, 3 Science Drive 2, 117543 Singapore}
\date{\today}

\begin{abstract}
Uncertainty relations state that there exist 
certain incompatible measurements, to which the outcomes cannot be 
simultaneously predicted. While the exact incompatibility of quantum 
measurements dictated by such uncertainty relations 
can be inferred from the mathematical formalism of quantum 
theory, the question remains whether there is any more fundamental 
reason for the uncertainty relations to have this exact form. What, if 
any, would be the operational consequences if we were able to go beyond 
any of these uncertainty relations? We give a strong argument that justifies
uncertainty relations in quantum theory by showing that violating them 
implies that it is also possible to violate the second law of thermodynamics. 
More precisely, we show that violating the uncertainty relations in quantum mechanics leads to 
a thermodynamic cycle with positive net work gain, which is very unlikely to exist in nature.  
\end{abstract}

\maketitle

Many features commonly associated with quantum physics, such as the 
uncertainty principle~\cite{heisenberg27} or non-locality~\cite{bell} appear 
highly counter-intuitive at first sight.
The fact that quantum mechanics is more non-local than any classical 
theory~\cite{bell}, but yet more limited~\cite{tsirel:original,tsirel:separated} 
than what the no-signalling principle alone demands~\cite{PR,PR1,PR2}
has been the subject of much investigation~\cite{function,glance,s3:nonlocal,causality,OppenheimWehner}. 
Several reasons and principles were put forward that explain 
the origin of such quantum mechanical limits~\cite{s3:nonlocal,causality,OppenheimWehner}.

In~\cite{OppenheimWehner} it was shown that the amount of non-locality 
in quantum mechanics is indeed directly related
to another fundamental quantum mechanical limit, namely the uncertainty 
principle~\cite{heisenberg27}.
This forged a relation between two fundamental quantum mechanical concepts. 
We may however still 
ask why the uncertainty principle itself is not maybe
stronger or weaker than predicted by quantum physics? - and, what would 
happen if it was? 

Here we relate this question to the second law 
of thermodynamics. We show that any violation of uncertainty relations 
in quantum mechanics also leads to a violation of the 
second law. 

\section{Background}

To state our result, we need to explain three different concepts. First, 
we need some properties of generalized physical 
theories (see e.g.~\cite{barrett,leifer,hardy,dariano,howard:survey}). 
Second, we recall the concept of uncertainty relations, and finally the second law of thermodynamics. 

{\bf Physical theories} Whereas it is not hard to prove our result for quantum 
theory, we extend our result to some more general physical theories.
These are described by a probabilistic framework that makes the minimal 
assumptions that there are \emph{states} 
and \emph{measurements} which can be made on a physical system (see, 
e.g.,~\cite{teleportation,entropy}). 
Even for general theories, we denote 
a state as $\rho \in \Omega$, where $\Omega$ is a convex state space.
In quantum mechanics, $\rho$ is simply a density matrix.
The assumption that the state space is convex is thereby generally 
made~\cite{howard:survey} and says that
if we can prepare states $\rho_1$ and $\rho_2$, then the 
probabilistic mixture $\rho = \rho_1/2 + \rho_2/2$ prepared by 
by tossing a coin and preparing $\rho_1$ or $\rho_2$ with probability $1/2$ each is also an element of $\Omega$. 
A state is called \emph{pure} if it cannot be written as a convex combination of other states.
Measurements consist of linear functionals $e_j: \Omega \rightarrow [0,1]$ called \emph{effects}.
We call an effect $e_j$ \emph{pure} if it cannot be written as a 
positive linear combination of any other allowed effects.
Intuitively, each effect corresponds to a possible measurement outcome, where $p(e_j|\rho) = e_j(\rho)$
is the probability of obtaining ''outcome'' $e_j$ given the state $\rho$.
More precisely, a measurement is thus given by $\textbf{e} = \{e_j \mid \sum_j p(e_j|\rho)=1\}$. 
For quantum mechanics, we will simply label effects by measurement operators.
For example, a projective measurement in the eigenbasis $\{0_Z,1_Z\}$ of the Pauli $Z$
operator is denoted by $p(0_Z|\rho) = \tr(\proj{0_Z}\rho)$.
The assumption that effects are linear, i.e., $p(e_j|\rho)$ is 
linear in $\rho$, is essentially made for all probabilistic 
theories~\cite{howard:survey}
and says that when we prepared a probabilistic mixture of states the 
distribution of measurement
outcomes scales accordingly. 

{\bf Uncertainty relations} A modern way of quantifying 
uncertainty~\cite{deutsch83,maassen88} is by means of 
\emph{entropic uncertainty relations} (see~\cite{WehnerWinter} for a survey), 
or the closely related \emph{fine-grained uncertainty 
relations}~\cite{OppenheimWehner}. 
Here we will use the latter. As for our cycle we will 
only need two measurements with two outcomes, 
and each measurement is chosen with probability $1/2$. 
We state their definition only for this simple case.
Let $\eff = \{f_0,f_1\}$ and $\gee = \{g_0,g_1\}$ denote the 
two measurements with effects $f_{y_1}$ and $g_{y_2}$ respectively.
A fine-grained uncertainty relation for these measurements is a set of inequalities
\begin{align}
	\label{eq:fineGrained}
	\left\lbrace \forall \rho:\  \frac{1}{2}\left(p(f_{y_1}|\rho) + p(g_{y_2}|\rho)\right) \leq \zeta_{\vec{y}} \middle|
	\vec{y} \in \{0,1\}^2
 	\right\rbrace\,.
\end{align}
To see why this quantifies uncertainty, note that if $\zeta_{\vec{y}} < 1$ for 
some $\vec{y} = (y_1,y_2)$, then
we have that if the outcome is certain for one of the 
measurements (e.g., $p(f_{y_1}|\rho) = 1$) it is uncertain ($p(g_{y_2}|\rho) < 1$) for the other.
As an example from quantum mechanics, consider measurements in the $X = \{0_X,1_X\}$ 
and $Z = \{0_Z,1_Z\}$ eigenbases.~\footnote{We use the common convention of labelling
the $X$ and $Z$ eigenbases states as $\{\ket{+},\ket{-}\}$ and $\{\ket{0},\ket{1}\}$ respectively.} 
We then have for all pure quantum states $\rho$ 
\begin{align}\label{eq:quantumFG}
	\frac{1}{2}\left(p(0_X|\rho) + p(0_Z|\rho)\right) \leq \frac{1}{2}+ \frac{1}{2\sqrt{2}}\,.
\end{align}
The same relation holds for all other pairs of outcomes $(0_X,1_Z)$,$(1_X,0_Z)$ and $(1_X,1_Z)$.
Depending on $\vec{y}$, the eigenstates of either 
$(X+Z)/\sqrt{2}$ or $(X-Z)/\sqrt{2}$ saturate these inequalities.
A state that saturates a particular inequality is also called 
a \emph{maximally certain state}~\cite{OppenheimWehner}.

For any theory such as quantum mechanics in which there is a direct correspondence between  
\emph{states} and \emph{measurements} uncertainty relations can also be stated in terms 
of \emph{states} instead of measurements.
More precisely, uncertainty relations 
can be written in terms of states if pure effects and pure 
states are dual to each other in the sense that 
for any pure effect $f$ there exists a corresponding pure 
state $\rho_f$, and conversely for every pure state $\sigma$ an effect
$e_\sigma$ such that $p(f|\sigma) = p(e_\sigma|\rho_f)$. Here, 
we restrict ourselves to theories that exhibit such a duality.
This is often (but not always) assumed~\cite{howard:survey,entropy}.
As a quantum mechanical example, consider
the effect $f = 0_X$ and the state $\sigma = \proj{0}$. We then 
have $p(f|\sigma) = \tr(\proj{+}\sigma) = \tr(\proj{+}\proj{0}) = p(e_\sigma|\rho_f)$
with $\rho_f = \proj{+}$ and $e_{\sigma} = 0_Z$.

For measurements $\eff = \{f_0,f_1\}$ and $\gee = \{g_0,g_1\}$ consisting of 
pure effects, let $\{\rho_{f_0},\rho_{f_1}\}$ and $\{\rho_{g_0},\rho_{g_1}\}$
denote the corresponding dual states. The equations of~\eqref{eq:fineGrained} 
then take the dual form
\begin{align}\label{eq:stateUR}
	\forall \mbox{ pure effects } e:\ \frac{1}{2}\left(p(e|\rho_{f_{y_1}}) + p(e|\rho_{g_{y_2}})\right) \leq \zeta_{\vec{y}}\,.
\end{align}
For our quantum example of measuring in the $X$ and $Z$ eigenbasis, 
we have $\rho_{0_X} = \proj{+}$, $\rho_{1_X} = \proj{-}$,
$\rho_{0_Z} = \proj{0}$ and $\rho_{1_Z} = \proj{1}$. We then have that for all 
pure quantum effects $e$
\begin{align}
	\frac{1}{2}\left(p(e|\rho_{0_X}) + p(e|\rho_{1_Z})\right) \leq \frac{1}{2}+ \frac{1}{2\sqrt{2}}\,.
\end{align}
The same relation holds for all other pairs $(0_X,1_Z)$,$(1_X,0_Z)$ and $(1_X,1_Z)$.
Again, measurement effects from the eigenstates of either 
$(X+Z)/\sqrt{2}$ or $(X-Z)/\sqrt{2}$ saturate these inequalities.
In analogy, with maximally certain states we refer to effects that saturate 
the inequalities~\eqref{eq:stateUR} as 
\emph{maximally certain effects}.
From now on, we will always consider uncertainty relations in terms of~\emph{states}.

{\bf 2nd law} The second law of thermodynamics is usually stated in terms of 
entropies. One way to state it is to say that the entropy of an 
isolated system cannot decrease. These entropies can be defined for general 
physical theories even for systems which are not described by the quantum 
formalism~\cite{entropy,barnum:entropy,japanese:entropy} (see appendix).
However, for our 
case it will be sufficient to consider one operational consequence of the 
second law of thermodynamics~\cite{peres,demon}: 
there cannot exist a cyclic physical process with a net work gain over 
the cycle.

\section{Result}

Our main result is that if it was possible to violate the fine-grained 
uncertainty relations as predicted by quantum physics, then we could 
create a cycle with net work gain. This holds for \emph{any} two projective measurements with 
two outcomes on a qubit. 
By the results of~\cite{OppenheimWehner} which showed that the amount of non-locality is 
solely determined by the uncertainty relations of quantum mechanics and our ability to 
steer, our result extends to a link between the amount of non-locality and the second law of thermodynamics. 

In the following we focus on the quantum case, i.e., in 
the situation where all the properties except the uncertainty relations 
hold as for quantum theory. In the appendix, we extend our result to more 
general physical theories that satisfy 
certain assumptions. In essence, different forms of entropies coincide in quantum mechanics, but
can differ in more general theories~\cite{entropy}. This has consequences on 
whether a net work gain in our cycle is due to a
violation of uncertainty alone, or can also be understood as the closely related question of whether 
certain entropies can differ.

\begin{figure*}[]
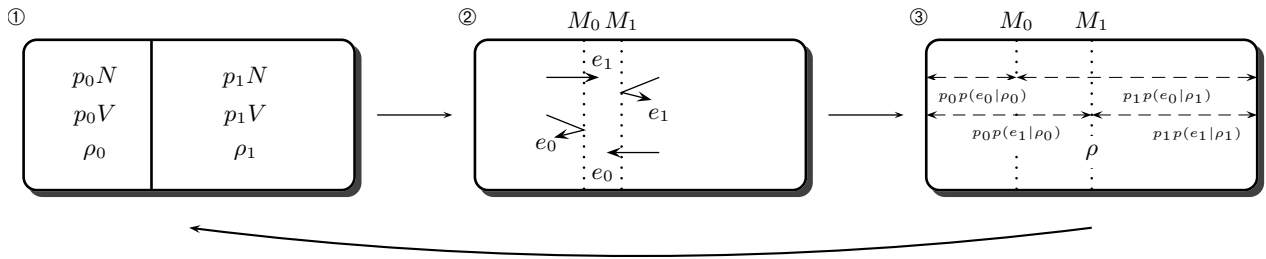

\centering
\pspicture*[](-8.4,-1.2)(8.4,2.5)
\rput[c]{0}(-6,0.0){
\rput[c]{0}(-2.3,2.3){\ding{192}}
\pspolygon[linewidth=1pt,linearc=5pt,shadow=true](-2.2,0)(2.2,0)(2.2,2)(-2.2,2)
\psline[linewidth=1pt]{-}(-0.5,0)(-0.5,2)
\rput[c]{0}(-1.25,1.5){$p_0N$}
\rput[c]{0}(-1.25,1){$p_0V$}
\rput[c]{0}(-1.25,0.5){$\rho_0$}
\rput[c]{0}(0.75,1.5){$p_1N$}
\rput[c]{0}(0.75,1){$p_1V$}
\rput[c]{0}(0.75,0.5){$\rho_1$}
}
\psline[linewidth=0.5pt]{->}(-3.50,1)(-2.5,1)
\rput[c]{0}(0,0.0){
\rput[c]{0}(-2.3,2.3){\ding{193}}
\pspolygon[linewidth=1pt,linearc=5pt,shadow=true](-2.2,0)(2.2,0)(2.2,2)(-2.2,2)
\psline[linewidth=1pt,linestyle=dotted]{-}(-0.75,0)(-0.75,2)
\psline[linewidth=1pt,linestyle=dotted]{-}(-0.25,0)(-0.25,2)
\rput[c]{0}(-0.75,2.25){$M_0$}
\rput[c]{0}(-0.25,2.25){$M_1$}
\psline[linewidth=0.5pt,arrowsize=4pt]{->}(-1.25,1)(-0.75,0.8)(-1.15,0.7)
\rput[c]{0}(-1.25,0.60){$e_0$}
\psline[linewidth=0.5pt,arrowsize=4pt]{->}(-1.25,1.5)(-0.55,1.5)
\rput[c]{0}(-0.5,1.70){$e_1$}
\psline[linewidth=0.5pt,arrowsize=4pt]{->}(0.25,1.5)(-0.25,1.3)(0.15,1.2)
\rput[c]{0}(0.25,1.0){$e_1$}
\psline[linewidth=0.5pt,arrowsize=4pt]{->}(0.25,0.5)(-0.45,0.5)
\rput[c]{0}(-0.5,0.20){$e_0$}
}
\psline[linewidth=0.5pt]{->}(2.50,1)(3.5,1)
\rput[c]{0}(6,0.0){
\pspolygon[linewidth=1pt,linearc=5pt,shadow=true](-2.2,0)(2.2,0)(2.2,2)(-2.2,2)
\rput[c]{0}(-2.3,2.3){\ding{194}}
\rput[c]{0}(-1,2.25){$M_0$}
\rput[c]{0}(-0.0,2.25){$M_1$}
\psline[linewidth=1pt,linestyle=dotted]{-}(-1.0,0)(-1.0,2)
\psline[linewidth=1pt,linestyle=dotted]{-}(0,0)(0,2)
\psline[linewidth=0.25pt,linestyle=dashed,arrowsize=3pt]{<->}(-2.2,1.5)(-1,1.5)
\rput[c]{0}(-1.45,1.25){\colorbox{white}{\tiny{$p_0p(e_0|\rho_0)$}}}
\psline[linewidth=0.25pt,linestyle=dashed,arrowsize=3pt]{<->}(-1.0,1.5)(2.2,1.5)
\rput[c]{0}(1.0,1.25){\colorbox{white}{\tiny{$p_1p(e_0|\rho_1)$}}}
\psline[linewidth=0.25pt,linestyle=dashed,arrowsize=3pt]{<->}(-2.2,1.0)(0,1.0)
\rput[c]{0}(-1.0,0.755){\colorbox{white}{\tiny{$p_0p(e_1|\rho_0)$}}}
\psline[linewidth=0.25pt,linestyle=dashed,arrowsize=3pt]{<->}(0.0,1.0)(2.2,1.0)
\rput[c]{0}(1.4,0.75){\colorbox{white}{\tiny{$p_1p(e_1|\rho_1)$}}}
\rput[c]{0}(0,0.5){\colorbox{white}{$\rho$}}
}
\psbezier{<-}(-6,-0.5)(-2,-1)(2,-1)(6,-0.5)
\endpspicture
\caption{The impossible process.}
\label{fig:imp}
\end{figure*}

Let us now first state our result for quantum mechanics more precisely.
We consider the following process as depicted in Figure~\ref{fig:imp}. We start with a box which contains two 
types of particles described by states $\rho_0$ and $\rho_1$ in two separated volumes. The state $\rho_0$ 
is the equal mixture of the eigenstates $\rho_{f_0}$ and $\rho_{g_0}$ of two measurements (observables)
$\textbf{f} = \{f_0,f_1\}$ and $\textbf{g} = \{g_0,g_1\}$. The state
$\rho_1$ is the equal mixture of $\rho_{f_1}$ and $\rho_{g_1}$. 
We choose the measurements such that the equal mixture $\rho = (\rho_0 + \rho_1)/2$ 
is the completely mixed state in dimension $2$. We 
then replace the wall separating $\rho_0$ from $\rho_1$ by two semi-transparent membranes, i.e., membranes which 
measure any arriving particle in a certain basis $\textbf{e} = \{e_0,e_1\}$ and only let it pass 
for a certain outcome. In the first part of the cycle we separate the two membranes until they are in 
equilibrium, which happens when the state everywhere in the box can 
be described as $\rho$. Then, in the second part of 
the cycle, we separate $\rho$ again into its different 
components. 

We find that the total work which can be extracted by 
performing this cycle is given by 
\begin{align}
\nonumber  \Delta W &=
  NkT \ln2 \left(
\sum_{i=0}^1  p_i S(\rho_i)
\right. \\ \nonumber & \quad \left.
-\frac{1}{2}  H\left(\zeta_{(f_0,g_0)}\right)
-\frac{1}{2}  H\left(\zeta_{(f_1,g_1)}\right)
 \right)\,.
\end{align}
Here, $S(\rho) = - \tr(\rho \log \rho)$ is the \emph{von Neumann entropy} of the state.
The entropy $H$ appearing in the above expression is simply the Shannon 
entropy of the distribution over measurement outcomes when measuring in the 
basis $\textbf{f}$ and $\textbf{g}$, respectively.~\footnote{The Shannon 
entropy of a probability distribution $\{p_1,\ldots,p_d\}$ is given 
by $H(\{p_1,\ldots,p_d\}) = - \sum_j p_j \log p_j$. All logarithms in this paper are to base $2$.} 

{\bf Example} To illustrate our result, consider the concrete quantum example, where 
the states are given by
\begin{align} \label{eq:stateDef}
\rho_0 &=
\frac{1}{2}\left( \rho_{0_X}+ \rho_{0_Z} \right)=\frac{\id + \frac{X+Z}{2}}{2}\ \text{and}\\
\nonumber \rho_1&=\frac{1}{2}\left( \rho_{1_{X}}+ \rho_{1_{Z}} \right)=
\frac{\id - \frac{X+Z}{2}}{2} \,.
\end{align}
The work which can be extracted from the 
cycle then becomes
\begin{align}
\nonumber  \Delta W &=
  NkT \ln2 \left(
H\left( \frac{1}{2}+\frac{1}{2\sqrt{2}}\right)
\right. \\ \nonumber & \quad \left.
-\frac{1}{2}  H\left(\zeta_{(0_X,0_Z)}\right)
-\frac{1}{2}  H\left(\zeta_{(1_X,1_Z)}\right)
 \right)\,.
\end{align}
The fine-grained uncertainty relations predict in the quantum case that 
$\zeta_{(0_X,0_Z)}$ and $\zeta_{(1_X,1_Z)}$ are at most 
$\frac{1}{2}+\frac{1}{2\sqrt{2}}$. We see that a theory which can violate 
this uncertainty relation, i.e., reach a larger value of $\zeta$, would 
lead to  $\Delta W>0$ --- a violation of the second law of thermodynamics.

\section{Methods}

We now explain in more detail how we obtain the work which can be 
extracted from the cycle in quantum mechanics. In the appendix, we consider
the case of general physical theories.

\subsection*{First part of the cycle}

For the first part of the cycle we start with two separate 
parts of the box in each of which there are $N/2$ particles in the 
states $\rho_0$ and $\rho_1$ respectively. These states 
are described by 
\begin{align} 
\nonumber \rho_0 &=
\frac{1}{2}\left( \rho_{f_0}+ \rho_{g_0} \right)\ \text{and}\\
\nonumber \rho_1&=\frac{1}{2}\left( \rho_{f_1}+ \rho_{g_1} \right) \,,
\end{align}
where $\textbf{f} = \{f_0,f_1\}$ and 
$\textbf{g} = \{g_0,g_1\}$ are chosen such that 
the state $\rho=\rho_0/2+\rho_1/2$ corresponds to the completely mixed state in dimension $2$. 
We then make a projective measurement $\textbf{e}=\{e_0,e_1\}$ with two possible outcomes denoted by $0,1$.
More precisely, we insert two semi-transparent membranes instead of the wall 
separating the two volumes. One of the membranes is transparent to $e_0$ but completely opaque to 
$e_1$ while the other lets the particle pass if the outcome is $e_1$, but not if it was 
$e_0$. Letting these membranes move apart until they are in equilibrium, we can 
extract work from the system. The equilibrium is reached when on both sides of the membranes which is 
opaque for $e_1$, there is the same density of particles in this state and similarly for the 
membrane which is opaque for $e_0$.   

The work which can be extracted from the first part of the 
cycle (i.e., by going from {\ding{192}} to \ding{194} in Figure~\ref{fig:imp}) is 
given by the following (see appendix). 
\begin{align}
\nonumber W &= 
 NkT \ln 2\left( 1
- \frac{1}{2}  H\left(\frac{1}{2}p(e_0|\rho_{f_0})
+ \frac{1}{2}p(e_0|\rho_{g_0}) 
\right)
\right. \\ \nonumber & \quad \left.
- \frac{1}{2}  H\left(\frac{1}{2}p(e_1|\rho_{f_1})+ \frac{1}{2}p(e_1|\rho_{g_1}) \right)
\right)\\
\nonumber & \leq 
NkT \ln 2\left( 1
- \frac{1}{2}  H\left(\zeta_{(f_0,g_0)}\right)
- \frac{1}{2}  H\left(\zeta_{(f_1,g_1)}\right)
\right)\,,
\end{align}
where we denoted by $\zeta$ the fine-grained uncertainty 
relations. The inequality can be saturated by choosing
$e_0$ and $e_1$ to be maximally certain effects.~\footnote{It is easy to see 
that in quantum mechanics the maximally certain effects $e_0$ and $e_1$ do 
indeed form a complete measurement in dimension $2$.} 
Note that our argument is not specific to the 
outcome combination $(0_f,0_g)$ and $(1_f,1_g)$ used in the the fine-grained uncertainty relation 
and choosing the remaining two inequalities corresponding to outcomes $(0_f,1_g)$ and $(1_f,0_g)$ 
leads to an analogous argument.

\textbf{Example} For our quantum example given by the states~\eqref{eq:stateDef}
we obtain 
\begin{align}
\nonumber W  & \leq 
NkT \ln 2\left( 1
- \frac{1}{2}  H\left(\zeta_{(0_X,0_Z)}\right)
- \frac{1}{2}  H\left(\zeta_{(1_X,1_Z)}\right)
\right)\,.
\end{align}
Equality is attained by taking $\{e_0,e_1\}$ to be the maximally certain effects
given by the two eigenstates of $(X+Z)/\sqrt{2}$.

\subsection*{Second part of the cycle}

In the second part we form a cycle (i.e., we go from {\ding{194}} to \ding{192} in Figure~\ref{fig:imp})). 
We start with the 
completely mixed state $\rho$. Denote the different pure components of 
$\rho$ by $\{q_j,\sigma_j\}_j$, i.e., $\rho=\sum_j q_j\sigma_j$. We 
can now  `decompose' $\rho$ into its components 
by inserting a semi-transparent membrane which 
is opaque for a specific component $\sigma_j$, but completely 
transparent for all other components. Effectively, this membrane 
measures using the effects $h_{\sigma_j}$ that
are dual to the states $\sigma_j$. This membrane is used 
to confine all states $\sigma_j$ in a volume $q_jV$. This is done for 
all components and we end up with a box where each component of $\rho$ is sorted 
in a volume proportional to its weight in the convex combination. 
This process needs work proportional to $S(\rho)$. 

In a second step, we create the (pure) components $\tau$ of 
$\rho_0=\sum_j r_j^0 \tau_j^0$ and $\rho_1=\sum_j r_j^1 \tau_j^1$
from the pure components of $\rho$ and then `reassamble' the states 
$\rho_0$ and $\rho_1$.  
In order to do so, we subdivide the volumes containing $\sigma_j$ 
into smaller volumes, 
such that the number of particles contained in these smaller 
volumes are proportional to $p_0r_j^0$ and $p_1r_j^1$. The pure 
state contained in each small volume is then transformed into the
pure state $\tau_j^0$ or $\tau_j^1$. Since these last states are 
also pure, no work is needed for this transformation. Finally, 
we `mix' the different components of $\rho_0$ together, which 
allows us to extract work $p_0S(\rho_0)$. Similarly we obtain 
work $p_1S(\rho_1)$ from $\rho_1$. 

In total, the transformation  
$\rho \rightarrow \{p_i,\rho_i\}$, needs work 
\begin{align}
\nonumber W&=NkT \ln 2 (S(\rho)-\sum_i p_i S(\rho_i))\,.
\end{align} 

{\bf Example} Returning to 
the example above and using that the two eigenvalues of $\rho$ 
are $1/2$, we obtain
\begin{align}
\nonumber S(\rho)&= -2\cdot \frac{1}{2} \log_2 \frac{1}{2}=1\,.
\end{align}
Both $\rho_0$ and $\rho_1$ have the two eigenvalues 
$\lbrace \frac{1}{2}+\frac{1}{2\sqrt{2}}, \frac{1}{2}-\frac{1}{2\sqrt{2}}\rbrace$. 
Therefore, 
\begin{align}
\nonumber S(\rho_i)&=  H\left( \frac{1}{2}+\frac{1}{2\sqrt{2}} \right)\approx H(0.85)\,.
\end{align}
The total work which has to be invested for this process is therefore given by 
\begin{align}
\nonumber W&=NkT \ln2\left( 1- H\left( \frac{1}{2}+\frac{1}{2\sqrt{2}} \right)\right) \,.
\end{align}

\subsection*{Closing the cycle} 

If we now perform the first and second process described above 
one after another (i.e., we perform a cycle, as 
depicted in Figure~\ref{fig:imp}), the total work which can be 
extracted is given by 
\begin{align}
\nonumber \Delta W &= NkT \ln2 \left( 
 - \left( S(\rho)- \sum_i p_i S(\rho_i) \right) 
\right. \\ \nonumber & \quad \left.
+
\left( 1
- \frac{1}{2}  H\left(\zeta_{(f_0,g_0)}\right)
- \frac{1}{2}  H\left(\zeta_{(f_1,g_1)}\right)\right) 
  \right)\,.
\end{align}
In general, we can see that when the uncertainty relation is
violated, this quantity can become positive and a positive 
$\Delta W$ corresponds to a violation of the second law of thermodynamics. 

{\bf Example} In our example, the above quantity 
corresponds to 
\begin{align}
\nonumber \Delta W &= 
  NkT \ln2 \left(
H\left( \frac{1}{2}+\frac{1}{2\sqrt{2}}\right)  
\right. \\ \nonumber & \quad \left.
-\frac{1}{2}  H\left(\zeta_{(0_X,0_Z)}\right)
-\frac{1}{2}  H\left(\zeta_{(1_X,1_Z)}\right)
 \right)\,.
\end{align}
The fine-grained uncertainty relations for quantum mechanics state that 
$\zeta_{(0_X,0_Z)},\zeta_{(1_X,1_Z)}\leq \frac{1}{2}+\frac{1}{2\sqrt{2}}$. When 
this value is reached with equality, then $\Delta W=0$ in the above calculation. 

On the other hand if these values were larger, i.e., the uncertainty relation could be 
violated, then the binary entropy of them would be smaller and $\Delta W$ becomes 
positive. 

\section{Discussion}

We give a strong argument why quantum mechanical uncertainty relations 
should not be violated. Indeed, as we show, 
a violation of the uncertainty relations would lead to an `impossible 
machine' which could extract net work from a cycle. 
Our result extends to more general theories than quantum theory - however, 
raises the question of which general form of entropy~\cite{entropy}
is most significant. In quantum mechanics, the different entropies of~\cite{entropy} 
coincide, meaning that if a physical theory is just like quantum mechanics, 
but with a different amount of uncertainty, net work can be extracted.

Our cycle is similar to the ones given in~\cite{peres,cost,demon}, which study related questions:
We can understand uncertainty relations as given in~\eqref{eq:fineGrained} as imposing a limit
on how well one of several bits of information can be extracted 
from a qubit using the given measurements~\cite{OppenheimWehner}. This means 
that the amount of uncertainty for all pairs of measurements that we could 
make directly imposes a limit on how much classical information we can 
store in each qubit. Indeed, in any system that is finite dimensional 
(possibly due to an energetic constraint), it is thus clear that the 
mere fact that we can only store a finite 
amount of information in a finite dimensional system (Holevo's 
bound~\cite{holevo}) demands that non-commuting measurements obey 
uncertainty relations. This shows that our example is closely related to 
the ones given in~\cite{plenio,pleniovitelli,cost,demon} where it 
has been shown that if it was possible to encode 
more than one bit of information in a qubit and therefore to violate the 
Holevo bound~\cite{holevo}, then it was also possible to violate the second law of thermodynamics. 

In~\cite{peres} similar consequences had been shown if one was able to perfectly 
distinguish non-orthogonal quantum states. The possibility of distinguishing non-orthogonal states 
is again directly related to the question of how much information we can store in a quantum state.

In future work, it might be interesting to investigate whether an 
implication also holds in the other direction. Does any violation of 
the second law lead to a violation of the uncertainty relations? 

We have investigated the relation between uncertainty and the 
second law of thermodynamics. A concept related to uncertainty is the 
one of complementarity. It is an open question, whether a 
violation of complementarity could also be used to build such an 
impossible machine.

\textbf{Acknowledgments:} We thank Christian Gogolin, 
Markus M\"uller, and Jonathan Oppenheim for helpful 
discussions and comments on an earlier draft.
EH and SW acknowledge support from the National Research 
Foundation (Singapore), and the
Ministry of Education (Singapore).

\bibliographystyle{apsrev}

\appendix

\section*{General theories}

In this appendix, we extend our result to more general physical theories. To this end, 
we need to introduce several additional assumptions and entropies. 
As quantum mechanics satisfies all 
assumptions made here, the derivation below can also be taken as a 
detailed explanation of the results claimed in the
front matter.

In the main text, we have shown that a violation of uncertainty relations 
leads to a violation of the second law of thermodynamics for the 
quantum case. More precisely, we have assumed that all processes can be 
described by the quantum formalism, with the exception of the 
uncertainty relations. We now want to show that our result still 
holds when the physical processes have to be described by a general 
convex theory. We need several assumptions on these theories, which 
we clearly state below. 

\subsection*{General assumptions}

As already outlined we will make three very common assumptions on 
a generalized physical theory. The first two are thereby essentially 
made everywhere~\cite{howard:survey}, the third is made very 
often (but not always (see e.g.~\cite{entropy}). We label assumptions as $A\cdot$.
Whereas such assumptions may seem rather elaborate, there are physical reasons for assuming them.
For example, a property known as bit symmetry~\cite{markusprl} implies A3, A6 and A7.

\begin{description}
\item[A1] The state space $\Omega$ is convex.
\item[A2\label{it:lin}] Effects are \emph{linear} functionals. 
\item[A3\label{it:dual}] Pure states are dual to pure effects as outlined in the 
background section. Uncertainty relations can thus be stated 
equivalently in terms of states or measurements\footnote{In~\cite{markusprl} it 
was shown that such a duality holds at least for any theory which has 
a property called `bit symmetry', which means that it allows for reversible 
computation.}. 
\end{description}
Next, we will assume that pure effects are \emph{projective} in that there exists a way to implement
them in a physical measurement such that if we repeatedly apply 
an effect $e_i$ by repeating the measurement we again obtain 
the same outcome. That is, $p(e_ie_i|\rho)=p(e_i|\rho)$ for all $\rho$,
where with some abuse of notation we take $p(e_ie_i|\rho)$ to be the probability 
of observing $e_i$ when making the measurement twice in a row, and don't consider
the outcome of the first. A measurement is projective if it consists only of projective effects. 
Note that this is not the same as demanding that post-measurement 
states are the same for all $\rho$, which has significant 
consequences~\cite{corsin:mthesis}.
\begin{description}
\item[A4] Pure effects are projective.
\end{description}
We will also assume that the unit effect $u$, i.e. 
the effect satisfying $u(\rho) = 1$ for all $\rho$, has a dual state that
is analogous to the maximally mixed state. If we accept duality between 
states and measurements, than this assumption is very natural.
\begin{description}
\item[A5\label{it:halfe}] If $f_0 + f_1 = u$ for two effects $f_0$ and $f_1$, 
then the dual states $\rho_{f_0} + \rho_{f_1} = \rho_{u}$
and $e(\rho_{u}/2) = 1/2$ for all pure effects $e$.
\end{description}
Note that this assumption again implies that we are dealing with the analogue of a qubit, i.e. a two-level 
system. It is possible to extend our statements for quantum mechanics to traceless two-outcome observables
but as this requires additional assumptions in generality, we omit it. 
Our next assumption, however, is rather strong and significant, and extends 
beyond the duality of states and measurements. It is of course satisfied by quantum mechanics.  
\begin{description}
\item[A6\label{it:cycle2}] Let $\rho = \sum_{j=1}^d q_j \sigma_j$ be a decomposition of $\rho$ into perfectly distinguishable pure states $\sigma_j$: 
Let $h_{\sigma_j}$ denote the pure effect dual to $\sigma_j$. Then $\sum_{j=1}^d h_{\sigma_j} = u$
and $h_k(\sigma_j) = \delta_{jk}$ for all $j$ and $k$.
\end{description}
Finally, we will also need that pure states can be transformed into 
different pure states and that this does not require any work. In quantum 
mechanics, this is justified since the transformation just corresponds 
to a unitary. 
\begin{description}
\item[A7\label{it:pure}] Let $\rho$ and $\sigma$ be pure states. Then 
the transformation $\rho\rightarrow \sigma$ is reversible (and thus does not require 
any work, neither can any work be gained from it). 
\end{description}

\subsection*{Entropies}

Several definitions of entropy are possible in generalized 
theories~\cite{entropy} that happen to coincide in quantum 
mechanics. The first is the so-called \emph{decomposition entropy} given as
\begin{align}
S(\rho) = \min_{\substack{\{p_j,\rho_j\}_j\\\rho = \sum_{j=1}^d p_j \sigma_j}} H(\{p_1,\ldots,p_d\})\,,
\end{align}
where the minimization is taken over decompositions into pure 
states $\sigma_j$ and $H$ is the Shannon entropy. Here, we will take the minimum
over decompositions into perfectly distinguishable pure states. Note that the resulting quantity
is equally well defined, but avoids an unnecessary strengthening of assumption A6.
To define the other notion of entropy, we will need the following definition of
\emph{maximally fine-grained measurements}~\cite{entropy} as 
measurements such that each of its effects cannot be re-expressed
as a non-trivial linear combination, i.e., 
\begin{align}
\nonumber \textbf{e}&=\{e_i\}_i :\text{ maximally fine-grained }\Leftrightarrow \\
\nonumber 
& \text{ for all }e_i:\ e_i=\alpha e_\alpha^\prime+\beta e_\beta^\prime,\ \alpha,\beta>0 \ \Rightarrow\ e_\alpha^\prime=e_\beta^\prime\,.
\end{align}
We also call any effect satisfying this equation fine-grained.
Note that pure effects are automatically maximally fine-grained. 
The \emph{measurement entropy} is then given by
\begin{align}
H(\rho) = \min_{\{e_j\}_{j=1}^\ell} H(\{e_1(\rho),\ldots,e_{\ell}(\rho)\})\,,
\end{align}
where the minimization is taken over maximally fine-grained measurements. 
Finally, it would be possible to define entropies by a thermodynamical process itself~\cite{vonNeumann:entropyDef}, even
in general physical theories~\cite{barrett:entropy}.
In such a setting, also a difference between the decomposition and measurement entropy can 
lead to a violation of the second law~\cite{jonathan:entropy}. As such, it is still under investigation what is the most relevant
entropy~\cite{entropy} in general theories, also when it comes to operational tasks from quantum information such as decoupling~\cite{markus:decoupling,markus:newdecoupling}.

\subsection*{First part of the cycle}

Below we state the explicit calculation of the work which can be 
extracted from the first part of the cycle. The 
measurements $\textbf{f} = \{f_0,f_1\}$ and $\textbf{g} = \{g_0,g_1\}$ are 
now not necessarily quantum mechanical, but do obey the assumptions above. 
We use a generalized notion of the completely mixed state and a projective measurement. 
In order to determine the position of the semi-transparent membranes 
in equilibrium, we assume that they perform a 
projective measurement, in the sense that a particle which is once 
measured to be $e_0$ ($e_1$) will give outcome $e_0$ ($e_1$) with certainty when 
the measurement is repeated, and never outcome $e_1$ ($e_0$). 
Additionally, we will use the following definitions in our calculation. 
\begin{description}
\item[D1\label{it:def}]  $\rho_0$ is the mixture of $\rho_{f_0}$ and $\rho_{g_0}$, and $\rho_1$ of $\rho_{f_1}$ and $\rho_{g_1}$, i.e., 
\begin{align} 
\nonumber \rho_0 &=
\frac{1}{2}\left( \rho_{f_0}+ \rho_{g_0} \right)\\
\nonumber \rho_1&=\frac{1}{2}\left( \rho_{f_1}+ \rho_{g_1} \right) \,.
\end{align}
\item[D2\label{it:halfp}] We choose an equal mixture of $\rho_0$ and $\rho_1$, i.e., $p_i=1/2$ for all $i$.
\end{description}
Note 
that by assumption A5 the state $\rho=\rho_0/2+\rho_1/2$ has analogous properties to the completely mixed state in dimension $2$, i.e., $p(e_j)=\sum_i p_i p(e_j|\rho_i)=1/2$ for all~$j$.
\begin{description}
\item[D3\label{it:binary}]  We make a measurement with binary outcomes, i.e., $p(e_j|\rho_i)=1-p(e_{\bar{j}}|\rho_i)$.
\end{description}
The numbers on top of the equation refer to the assumptions and/or definitions
stated above which are used in this step of the calculation. 
\begin{align}
\nonumber W &=  NkT\left(
\sum_{i,j}
p_i p(e_j|\rho_i)\ln(p_i p(e_j|\rho_i))
\right. \\ \nonumber & \quad \left.
-
\sum_j p(e_j)\ln p(e_j) - \sum_i p_i \ln p_i \right) \\
\nonumber 
&\mathop{=}^{\ref{it:halfp},\ref{it:halfe}} NkT \ln 2\left(
\sum_{i,j}
p_i  p(e_j|\rho_i)\log (p_i p(e_j|\rho_i))
\right. \\ \nonumber & \quad \left.
-
\log \frac{1}{2} -  \log \frac{1}{2}\right) \\
\nonumber 
&\mathop{=}^{\ref{it:halfp}} 
\nonumber 
NkT \ln 2\left( 2
\vphantom{\frac{1}{2}}\right. \\ \nonumber & \quad \left.
+
\frac{1}{2}  \sum_{i,j}
p(e_j|\rho_i)\left ( \log \frac{1}{2}+\log  p(e_j|\rho_i))\right) \right)
\\
&= 
\nonumber 
NkT \ln 2\left( 1+
\frac{1}{2}  \sum_{i,j}
p(e_j|\rho_i)\left ( \log  p(e_j|\rho_i))\right) \right)
\\
&\mathop{=}^{\ref{it:binary}} 
\nonumber 
NkT \ln 2\left( 1+
\frac{1}{2}  \sum_{i}\left( 
p(e_j|\rho_i) \log  p(e_j|\rho_i)
\right. \right. \\ \nonumber & \quad \left.  \vphantom{\frac{1}{2}}  \left.
+
(1-p(e_j|\rho_i)) \log (1- p(e_j|\rho_i))
\right)
\right)
\\
&= 
\nonumber 
NkT \ln 2\left( 1
- \frac{1}{2}  \sum_{i}H(p(e_j|\rho_i))
\right)
\\
&= 
\nonumber 
NkT \ln 2\left( 1
- \frac{1}{2} H\left(p\left(e_j\middle|\frac{1}{2}\rho_{f_0}+\frac{1}{2}\rho_{g_0} \right)\right)
\right. \\ \nonumber & \quad \left.
-
\frac{1}{2} H\left(p\left(e_j \middle| \frac{1}{2}\rho_{f_1}+\frac{1}{2}\rho_{g_1}\right)
\right)
\right)\\
&\mathop{=}^{\ref{it:lin}} 
\nonumber 
NkT \ln 2\left( 1
- \frac{1}{2} H\left(\frac{1}{2} p\left(e_j\middle |\rho_{f_0}\right)+\frac{1}{2}p\left(e_j\middle|\rho_{g_0}\right)\right)
\right. \\ \nonumber & \quad \left.
-
\frac{1}{2} H\left(p\left(e_j\middle|\rho_{f_1}\right)+\frac{1}{2}p\left(e_j\middle|\rho_{g_1}\right)\right)
\right)\\
\nonumber 
&\mathop{=}^{\ref{it:dual}}  NkT \ln 2\left( 1
- \frac{1}{2}  H\left(\frac{1}{2}p(f_0|\rho_{e_j})
+ \frac{1}{2}p(g_0|\rho_{e_j}) 
\right)
\right. \\ \nonumber & \quad \left.
- \frac{1}{2}  H\left(\frac{1}{2}p(f_1|\rho_{e_j})+ \frac{1}{2}p(g_1|\rho_{e_j}) \right)
\right)\\
\label{eq:cycle1} & \leq 
NkT \ln 2\left( 1
- \frac{1}{2}  H\left(\zeta_{(f_0,g_0)}\right)
- \frac{1}{2}  H\left(\zeta_{(f_1,g_1)}\right)
\right)\,.
\end{align}

Equality is achieved when the measurement can be formed from the maximally certain effects.

\subsection*{Second part of the cycle}

We calculate the work needed for the second part of the 
cycle in two parts. First let us calculate the work needed to `decompose' $\rho$ into 
its different pure components. We use the effects $h_{\sigma_j}$, which 
are dual to the pure states $\sigma_j$ which form the components of 
$\rho$, i.e., $\rho=\sum_j q_j \sigma_j$. Note that we can do this for any decomposition, in particular the one minimizing $S(\rho)$.
\begin{align}
\nonumber W&=-NkT \ln 2 \left(\sum_j p(h_{\sigma_j}|\rho)\log p(h_{\sigma_j}|\rho) \right)\\
 &= 
 \nonumber
 -NkT \ln 2 \left(\sum_j p\left(h_{\sigma_j}\middle|\sum_{j^\prime}q_{j^\prime}\sigma_{j^\prime}\right)
  \right. \\ & \quad \left. 
   \nonumber
  \log p\left(h_{\sigma_j}\middle|\sum_{j^\prime}q_{j^\prime}\sigma_{j^\prime}\right) \right)\\
 &\mathop{=}^{\ref{it:lin}} 
\nonumber
 -NkT \ln 2 \left(\sum_j \left(\sum_{j^\prime} q_{j^\prime} p\left(h_{\sigma_j}\middle|\sigma_{j^\prime}\right)\right)
 \right. \\ & \quad \left. 
  \nonumber
 \log \left(\sum_{j^\prime} q_{j^\prime} p\left(h_{\sigma_j}\middle|\sigma_{j^\prime}\right)\right) \right)\\
  &\mathop{=}^{\ref{it:cycle2}}
  \nonumber
   -NkT \ln 2 \left(\sum_{j^\prime} q_{j^\prime} \log q_{j^\prime}  \right)\\
\label{eq:decompentropy}
   &=  NkT \ln 2 S(\rho)\,.
\end{align} 
We then transform the pure states $\sigma_j$ into the pure states $\tau^0_j$ or $\tau^1_j$. 
By~\ref{it:pure}, this does not require any work. 
By performing a processes analogous to~\eqref{eq:decompentropy} but in the reverse direction, 
we can then extract work $NkT \ln 2 \sum_i p_i S(\rho_i)$ by 
`reassembling' the states $\rho_0$ and $\rho_1$. Overall, the work 
needed for the second part of the cycle is given by
\begin{align}
 W&=NkT \ln 2 (S(\rho)-\sum_i p_i S(\rho_i))\,.
\label{eq:cycle2}
\end{align}

\subsection*{Closing the cycle} 

From the above calculation, i.e., by substracting~\eqref{eq:cycle2} 
from~\eqref{eq:cycle1}, we see that for the total 
cycle, the amount of work which can be extracted is 
given by  
\begin{align}
\nonumber \Delta W &= NkT \ln2 \left( 
 - \left( S(\rho)- \sum_i p_i S(\rho_i) \right) 
\right. \\ \nonumber & \quad \left.
+
\left( 1
- \frac{1}{2}  H\left(\zeta_{(f_0,g_0)}\right)
- \frac{1}{2}  H\left(\zeta_{(f_1,g_1)}\right)\right) 
  \right)\,,
\end{align}
however, where $S$ is now the general decomposition 
entropy.  A net work gain of this cycle, and therefore 
a violation of the second law of thermodynamics, can 
therefore be reached if the uncertainty relations 
can be violated without at the same time changing 
the decomposition entropy.

\end{document}